\documentclass[conference]{IEEEtran}
\IEEEoverridecommandlockouts
\usepackage{cite}
\usepackage{amsmath,amssymb,amsfonts}
\usepackage{algorithmic}
\usepackage{graphicx}
\usepackage{textcomp}
\usepackage{xcolor}
\usepackage{listings}
\usepackage{url}
\usepackage{array}
\usepackage{tikz} 
\usepackage{balance}

\def\BibTeX{{\rm B\kern-.05em{\sc i\kern-.025em b}\kern-.08em
    T\kern-.1667em\lower.7ex\hbox{E}\kern-.125emX}}

\newcolumntype{P}[1]{>{\centering\arraybackslash}p{#1}}
    
\begin{document}

\title{Translating Natural Language Queries to SQL Using the T5 Model\\
\thanks{We acknowledge the support of the Natural Sciences and Engineering Research Council of Canada (NSERC), Harris SmartWorks Division of Harris Computers, Okanagan College, and Langara College.}
}

\makeatletter
\newcommand{\linebreakand}{%
 \end{@IEEEauthorhalign}
 \hfill\mbox{}\par
 \mbox{}\hfill\begin{@IEEEauthorhalign}
}
\makeatother

\author{
\IEEEauthorblockN{Albert Wong}
\IEEEauthorblockA{\textit{Mathematics and Statistics} \\
\textit{Langara College}\\
Vancouver, Canada \\
0000-0002-0669-4352}
\and
\IEEEauthorblockN{Lien Pham}
\IEEEauthorblockA{\textit{Mathematics and Statistics} \\
\textit{Langara College}\\
Vancouver, Canada \\
Email: honglien.pham@gmail.com}
\and
\IEEEauthorblockN{Young Lee}
\IEEEauthorblockA{\textit{Mathematics and Statistics} \\
\textit{Okanagan College}\\
Kelowna, Canada \\
Email: yolee3112@gmail.com}

\and
\IEEEauthorblockN{Shek Chan}
\IEEEauthorblockA{\textit{Mathematics and Statistics} \\
\textit{Langara College}\\
Vancouver, Canada \\
0000-0002-5932-5390}
\linebreakand 

\and
\IEEEauthorblockN{Razel Sadaya}
\IEEEauthorblockA{\textit{Mathematics and Statistics} \\
\textit{Langara College}\\
Vancouver, Canada \\
Email: razelsadaya@gmail.com}

\and
\IEEEauthorblockN{Youry Khmelevsky}
\IEEEauthorblockA{\textit{Computer Science} \\
\textit{Okanagan College}\\
Kelowna, Canada \\
0000-0002-6837-3490}

\and
\IEEEauthorblockN{Mathias Clement}
\IEEEauthorblockA{\textit{Computer Science} \\
\textit{Okanagan College}\\
Kelowna, Canada \\
0000-0001-8206-307X}

\and
\IEEEauthorblockN{Florence Wing Yau Cheng}
\IEEEauthorblockA{\textit{Mathematics and Statistics} \\
\textit{Langara College}\\
Vancouver, Canada \\
Email: fcheng18@mylangara.ca}

\linebreakand 

\and
\IEEEauthorblockN{Joe Mahony}
\IEEEauthorblockA{\textit{Research and Development} \\
\textit{Harris SmartWorks}\\
Ottawa, Canada \\
JMahony@harriscomputer.com}

\and
\IEEEauthorblockN{Michael Ferri}
\IEEEauthorblockA{\textit{Research and Development} \\
\textit{Harris SmartWorks}\\
Ottawa, Canada \\
mferri@harriscomputer.com}
}

\maketitle

\begin{abstract}

This paper presents the development process of a natural language to SQL model using the T5 model as the basis. The models, developed in August 2022 for an online transaction processing system and a data warehouse, have a 73\% and 84\% exact match accuracy respectively. These models, in conjunction with other work completed in the research project, were implemented for several companies and used successfully on a daily basis. The approach used in the model development could be implemented in a similar fashion for other database environments and with a more powerful pre-trained language model. \end{abstract}


\section{Introduction}

The increasing volume of data collected by utility companies due to the installation of automated metering infrastructure (AMI) \cite{RashedMohassel2014AInfrastructure} has intensified the need and interest in implementing solutions that provide effective data storage and retrieval that are synchronized \cite{Wong2021ASystems}. While recent advances in data warehousing and other data depository technologies provide ample choices in organizing and storing massive amounts of data, the query and retrieval of these data require the often expensive intervention of a software engineer. The ideal scenario remains that an end user could interact, through English or other natural languages, directly with the data depository to produce the desirable data insights.

A considerable amount of research has been completed in the last ten years to address the query problem. For the most part, a combination of semantic analysis and some form of neural network is used in the algorithm for the translation (from NL to SQL) problem. A literature review for work completed up to the summer of 2021 can be found in \cite{Wong2021ASystems, Affolter2019ADatabases}. With the rapid advance in deep learning and natural language processing, these reviews have quickly become obsolete as research in this area has intensified \cite{Li2023RESDSQL:Text-to-SQL, Li2023Graphix-T5:Parsing, Pourreza2023DIN-SQL:Self-Correction}. This is also applicable to the research study that we will discuss in detail here.

The research study presented here is a response to the needs of the utility industry as described above. Working with an industrial partner in the utility service industry, the project is a case study of how an end-to-end natural language data query system could be built quickly using available third-party components and basic hardware. 

It will become clear that the project needs to build some basic components: a user interface with voice and text input and graphical or tabular output, a high-performance data warehouse that allows for effective storage and speedy response, as well as an NL to SQL model that translates English queries into the corresponding executable SQL statement.  The first component is readily available and therefore its development and implementation is a matter of routine engineering. It is the second and third components that are the focus of the research study. 

The main contributions of the research study are the development of an effective data warehouse and the insights and ideas generated through the development process using a large pre-trained language model. Contributions made in the data warehouse arena were documented and published in \cite{Joiner2022DWStudy} and therefore will not be repeated here. Work completed in the NL to SQL modelling is documented in detail in the following sections.

This work could be used for developing future generations of natural language data query systems. As well, the experience gained in the integration of the three components in the development and implementation of a viable product could also be useful for the industry.

In the following, we present an overview of the research work completed in the area of converting natural language (NL) queries into SQL in Section II. Our research project is introduced in Section III. Research work on the NL to SQL modelling and the results are presented in detail in Section IV. Significant results published after the completion of our research efforts are briefly described in Section V. We close by highlighting the conclusions that we drew from the research efforts.

\section{Literature Review}

In this section, work on translating a natural English query to a database into the corresponding SQL query statement is reviewed. This concept is not new, and various attempts have been made. A comprehensive review of these efforts up to the year 2021 can be found in \cite{Wong2021ASystems}. 

There are three distinct approaches to converting an NL query into a database query: rule-based algorithms, ML algorithms \cite{Wong2021ASystems}, and the use of large pre-trained language models. In the following, a quick review of work completed at the time of this research is divided as follows: work using a rule-based algorithm is presented in the Rule-based Algorithm subsection. Methods using neural networks or a deep learning model are presented under the Machine Learning and Deep Learning Models subsection. Recent work that has been completed using a pre-trained language model is presented in the last subsection. Finally, a number of publicly available datasets used frequently to develop models in this area are presented to complete the review.

\subsubsection{Rule-based Algorithms}

Rule-based algorithms have three main components: a natural language Processing (NLP) processor, a mapping table, and the corresponding generation engine \cite{Wong2021ASystems}.  The NLP processor is not unique to rule-based algorithms. It is responsible for tokenization, stop words elimination, word replacement, and parsing which is also used in machine learning algorithms. 

Tokenization, stop word removal, stemming or lemmatization, and parsing are determined to enhance the speed of the matching process, diminish the ambiguity, and increase the accuracy of matching \cite{Wong2021ASystems}. Rule-based algorithms have been applied to smaller datasets that yielded good results as evidenced in Uma \cite{Uma2019FormationNLP}. 

Although rule-based algorithms can be used for text-to-SQL tasks, they are suitable for simple queries and databases\cite{Deng2022RecentExpect}. Developing rules for large, cross-domain datasets like Spider \cite{Yu2018Spider:Task} may require a huge amount of effort in generating the rules.

\subsubsection{Machine Learning Algorithms}

With the advancement of machine learning algorithms through the years, much work has been done using machine learning models to accomplish text-to-SQL generation tasks. Recurrent Neural Networks (RNN) and Graph Neural Networks (GNN) are the most common types of neural networks used in translating text to an SQL query. The long short-term memory (LSTM), gate recurrent neural network (GRU), sequence-to-sequence (seq2seq), and attention model are frequently used in RNN models \cite{Wong2021ASystems}.  

F-SemtoSQL \cite{Li2020AModel} and IRNet \cite{Guo2019ContentGeneration} models used two bidirectional LSTM combined with the application of an attention mechanism to encode a question sentence. Authors in [16] also used two bidirectional GRUs with an attention mechanism in the prediction step of their model. Seq2SQL \cite{Zhong2017Seq2SQL:Learning} used two-layer bidirectional LSTM while \cite{Mellah2021SQLNetworks} used both LSTM and GRU for this task. The authors in \cite{Bogin2019RepresentingParsing, Chen2021ShadowGNN:Parser, Hui2022S2SQL:Parsers,Cai2021SADGA:Text-to-SQL,Wang2019RAT-SQL:Parsers} utilized GNNs in their models to represent schema as a graph structure.

Neural language modelling typically use word embeddings such as Word2Vec or GloVe \cite{Pennington2014GloVe:Representation} as first layers \cite{Mellah2021SQLNetworks,Zhong2017Seq2SQL:Learning, Ferreira2020EvaluatingSystems,Vathsala2021NLP2SQLLearning,Xu2017SQLNet:Learning}. This layer provides association and similarity of words prior to training \cite{Wong2021ASystems}. These methods improve the performance of downstream NLP tasks but lack the ability to represent the contextual meaning of the words\cite{Wang2022Pre-TrainedApplications}.

Typical models that tackle text-to-SQL tasks come under an encoder-decoder scheme. Some researchers have shown that input and output adjustment can improve the accuracy of the model\cite{Wong2021ASystems}. Several researchers have utilized different methods during encoding to improve schema linking and schema encoding. Encoding token types \cite{Guo2019ContentGeneration,Brunner2021ValueNet:Information} such as table, column, or value can be used to represent the linkage between the question and the schema. Graph-based methods are also widely used to represent the rich structural information of database schemas \cite{Deng2022RecentExpect}. S2SQL \cite{Hui2022S2SQL:Parsers} used a relational graph attention network to leverage syntactic information of questions. Cai et al. proposed a unified encoding model SADGA \cite{Cai2021SADGA:Text-to-SQL} using graph structure to help in schema linking.  RAT-SQL \cite{Wang2019RAT-SQL:Parsers} utilized an attention model as well as a graph neural network to handle various pre-defined relations such as “both columns are from the same table” \cite{Deng2022RecentExpect}.

For decoding, sketch-based slot-filling methods \cite{Hwang2019AContextualization,Xu2017SQLNet:Learning,Choi2020RYANSQL:Databases} have been employed to generate an SQL query by using different modules to predict the content for each corresponding slot from a sketch. On the other hand, generation-based methods \cite{Guo2019TowardsRepresentation,Wang2019RAT-SQL:Parsers,Cao2021LGESQL:Relations} used abstract syntax trees to decode the SQL query. 

\subsubsection{The Use of Pre-trained Language Models}

Pre-trained language models are large neural networks that are trained over a large text corpus and can be fine-tuned on a downstream NLP task \cite{Elazar2021MeasuringModels}. Some researchers have used the pre-trained language model BERT \cite{Devlin2018BERT:Understanding} as the first layer in their machine learning models \cite{Hwang2019AContextualization,Guo2019TowardsRepresentation,Zhang2019Editing-BasedQuestions,Li2020AModel,Ma2020MentionGeneration,Cai2021SADGA:Text-to-SQL,Cao2021LGESQL:Relations,Wang2019RAT-SQL:Parsers,Choi2020RYANSQL:Databases,Lin2020BridgingParsing} to leverage contextual information.

The availability of large datasets such as Spider, WikiSQL, or SparC, has enabled researchers to fine-tune the model for text-to-SQL tasks. For example, Shaw et al. \cite{Shaw2020CompositionalBoth} showed competitive results from fine-tuning the Text-to-Text Transfer Transformer (T5) model \cite{Raffel2019ExploringTransformer} without relational structures. Authors of UnifiedSKG \cite{Xie2022UnifiedSKG:Models} achieved state-of-the-art results using the T5 model for various semantic parsing tasks including text-to-SQL.

Using pre-trained language models, PICARD \cite{Scholak2021PICARD:Models} attempted to constrain the auto-regressive decoder of language models through incremental parsing. This method can be operated directly on the output of a pre-trained language model such as T5. The authors claimed to have significantly improved performance on the Spider dataset. RASAT \cite{Qi2022RASAT:Text-to-SQL} also tried to improve the performance of pre-trained language models in text-to-SQL tasks by incorporating relational structures such as schema linking and schema encoding while still inheriting the pre-trained parameters from the T5 model effectively.

\subsubsection{Spider and Other Public Data Sets for Development}

A number of data sets, such as Spider, WikiSQL, and Sparc, have been compiled to support the development of the NL to SQL models and for establishing benchmarks in comparison of accuracy for these models. The availability of these data sets has also enabled researchers to fine-tune their models. We have chosen to use the Spider data set as part of the development efforts in this research.

\section{Building the NL to SQL Model for a Utility Service Company} 

 As mentioned, the increasing volume of data collected has driven the interest in finding solutions that provide effective and simplified data retrieval  \cite{Wong2021ASystems}. Hence, recent advances in the use of deep learning for natural language processing have gained attention for its application to text-to-SQL tasks that enable effective interfaces between natural language and database systems such as OLTP and data warehouses.

This research focused on the development of the capability of text-to-SQL models in retrieving information from the database systems for a leading provider of utility software and data solutions. Individual clients' utility usage is automatically, precisely, and frequently measured by ``smart meters", providing detailed and accurate data to a utility company. The vast amount of data generated by smart metering infrastructure for utilities has immense potential to transform utility operations, which can result in enhanced safety and dependability, better service to the public, and reduced utility expenses for businesses, homeowners, and tenants.

The use of a text-to-SQL solution aims to aid in easier information retrieval by taking non-technical, plain language questions like ‘Which location has the highest average annual electricity usage?’ and translating those questions into SQL queries against the OLTP or a data warehouse (DW).

\subsection{T5 Model}
In this paper, the approach we used for the text-to-SQL task is the training of the T5 model using a customized training data set. The T5 is a pre-trained, transformer-based model that aims to have a unified framework that transforms text in one format into another \cite{Raffel2019ExploringTransformer}. Every NLP task is cast as feeding the model with text as input and then training it to generate some text output. 

There are several variations of the T5 model. Our experiments used the T5-base as the base model for its smaller size. We trained the T5-base model using a set of custom-developed training data. Parameters such as the number of epochs, batch size, learning rate, etc. were fine-tuned using mainly the free Kaggle online computing resource. The project aims to demonstrate that an effective NL to SQL model that is sufficiently accurate could be rapidly developed using basic computing resources.  

\subsection{Training and Testing Datasets}
To fine tune the T5 model for the text-to-SQL downstream task, the large-scale, cross-domain, text-to-SQL Spider \cite{Yu2018Spider:Task} dataset was used. The Spider dataset used in our experiments has a total of 8,659 training queries and 1,034 testing queries for 166 databases. A natural language query to a database and its corresponding SQL query are the two main fields in this dataset.

In addition to the Spider dataset, we also used two sets of custom datasets associated with the utility service company's database systems. The first one is for the OLTP database which has a total of 1,705 queries where 1,192 of the queries are for training while 513 queries are for testing. The second custom dataset is for the data warehouse which has a total of 597 queries where 440 of the queries are for training and 157 queries are for testing. 

Again, a natural language query to a database and its corresponding SQL query are the two main fields in these datasets. Queries in the custom training datasets were developed so that all columns and tables in the OLTP or the data warehouse were covered at least a couple of times. This is to ensure that the T5 model is ``made aware" of them during the training process. Queries in the testing datasets were similarly developed to test whether the resulting model could translate the natural language queries that are often used in real situations and have coverage over the OLTP or the data warehouse. These custom datasets were developed iteratively and jointly with the technical staff from the utility service company and were the most time-consuming part of the research and development process. 

Two models were developed in the research; one for the OLTP and one for the DW. The training data sets used to train the T5 model are a concatenation of the Spider data set and the corresponding custom training dataset. The idea was that the Spider dataset was used to train the model in the ``grammar" of SQL translation while the custom dataset ensured that the model learned the necessary ``vocabulary" (name of columns and tables) so that the model could provide effective translation for queries from the natural language to SQL.

The development was completed on both Kaggle cloud computing environment using GPU P100 and on our own Ubuntu server with an NVIDIA GPU. A typical training run of the model takes about 4 hours.    

\subsection{SQL Correction}
After using the trained T5 model to translate the natural language query to SQL, we implemented a simple post-processing method to correct the SQL query with reference to the database schema. 

This process scans the output SQL for incorrect names of columns and tables and, if necessary, corrects them according to the database schema. For example, it will change “meter” in the SQL query to ``meters” if the correct name for the column in the schema is ``meters". It is done using a character-by-character, sequential match that is based on positions. It replaces tokens with the correct tables or column names based on the information from the database schema. 

\subsection{Performance Metrics}
For model evaluation, Zhong proposed two evaluation metrics from different perspectives: logical form accuracy and execution accuracy \cite{Zhong2017Seq2SQL:Learning}.

Logical form accuracy is defined as the percentage of generated queries that are converted correctly from the actual query. On the other hand, execution accuracy stands for the percentage of generated queries that can be executed against the database and produce the correct results.

Zhong mentioned that two database queries can produce the same result. If only the logical form accuracy is utilized, some generated queries with a correctly executed result but not in the same syntax would be treated as incorrect queries. It is suggested that both of the metrics should be considered to evaluate the performance of models \cite{Zhong2017Seq2SQL:Learning}.

For this research, we used the exact match accuracy with manual inspection of the testing results to measure performance. We believe that with manual inspection and adjustments, the exact match accuracy and execution accuracy are the same in this case.

Two datasets were used to evaluate performance: one used only the test datasets described above with queries not seen by the model (Test 1); the other the testing and the custom-developed training dataset together (Test 2). This was applied in the OLTP model as well as the data warehouse model. It was expected that the performance for the first test dataset would be lower and be a better measure of the accuracy of the model.

\subsection{Results and Discussions}
Using the training datasets as described above, the model was trained and tuned iteratively with adjustments made mainly to the training datasets. We evaluated the models using the two sets of test data as described above. For the OLTP queries, the final model achieved a 72.9\% exact match accuracy on Test 1 and an 83.7\% exact match accuracy on Test 2. On the other hand, the accuracy is higher for the data warehouse queries with 85.4\% exact match accuracy on Test 1 and 87.5\% exact match accuracy on Test 2.

The results for the data warehouse model were better in terms of exact match accuracy. This was expected as the data warehouse has a simpler database schema and the queries tend also to be simpler and therefore easier to ``learn".

Compared with other works that use the T5 model for text-to-SQL tasks, the performance of our model is similar or better than the results of previous experiments. For instance, PICARD \cite{Scholak2021PICARD:Models} has an exact-set match accuracy of 65.8\% on T5-base and 75.5\% on the larger T5-3B model on Spider’s development test set. RASAT \cite{Qi2022RASAT:Text-to-SQL}, which employs a more sophisticated architecture with T5, has an exact-set match accuracy of 72.6\% on the development test set. Lastly, UnifiedSKG \cite{Xie2022UnifiedSKG:Models} showed 58.12\% match accuracy on their model trained based on the T5-base model.

The development of the models was completed in August 2022. The two models, along with the chatbot developed by students in the Okanagan College research team and a graphical interface for output generated from the SQL queries, have since been implemented for several clients by the utility service company in early 2023. The models were proven to be usable based on the experience of using the models in real situations.

While the results of the research have been proven through implementation, they could be further improved by using a more powerful version of a pre-trained language model such as T5-Large, T5-3B, T5-11B or a more recent model. Also, while the approach employed allowed for a development path that is straightforward, research on how to generate the set of training queries quickly based on the database schema would further improve the speed of the development efforts and the deployment of the model to other database environments.

\subsection{Research Completed since the Completion of The Project}

After our model was completed in August 2022, some research groups developed new architectures based on T5, and these models further improved the accuracy of the text-to-SQL parser, taking advantage of database schema linkages. For example, Graphix-T5 \cite{Li2023Graphix-T5:Parsing} combined the standard T5 encoding layer with semi-pretrained relational graph neural network (GNN) to integrate semantic information and develop scheme linkage relationship, which achieved 74\% accuracy rate in exact match text-to-SQL evaluation. While most of the groups focus on how the NLP questions can generate the corresponding SQL, Zhao et al \cite{Zhao2022ImportanceParsing} improved the model from SQL to text direction (SQL-to-text), they identified the existing limitation of current synthesis methods and proposed the synthesis framework which used intermediate representation (IR) to preserve the important information from the query to construct the correspondence text of the natural language query (NLQ) as the parallel data. The synthesis framework achieved 73.1\% accuracy of an exact match. Zeng et al. \cite{Zeng2022N-BestSystems} improved the correctness and the coherence of query execution plan and the schema linkage, and reached 72.2\% exact match accuracy. 

These groups demonstrated that the linkages within the database schema could significantly improve the accuracy rate. Li et al. \cite{Li2023RESDSQL:Text-to-SQL} not only made use of the database schema linkages but also proposed the Ranking-enhanced Encoding plus a Skeleton-aware Decoding framework for Text-to-SQL (RESDSQL) framework, which enhances the encoding system with a ranking-based approach using the schema items. By the ranking approach, the schema items with less probability can be filtered therefore it can reduce the noise from the schemas. In the decoder, they first generated the SQL structure and then the complete SQL statement, which helps reduce the challenges associated with text-to-SQL conversion.

Another approach to enhance execution performance is by improving the training technique through self-play \cite{Liu2022AugmentingSelf-Play}. This method leverages interactions to generate and refine SQL queries based on multiple conversational turns. The self-play approach allows the model to learn from diverse query-generation scenarios through iterative iterations. Each iteration contributes to refining the SQL statements, leading to the development of more robust and accurate text-to-SQL models over time. This iterative refinement enables the model to use less training data as the process augments the data progressively. However, this training approach has its limitations. In the real world, when someone misunderstands a concept and subsequently makes decisions based on that misunderstanding, rectifying those decisions can be challenging.

In addition to the T5 model, several other models like GPT-4 and CodeX\cite{Pourreza2023DIN-SQL:Self-Correction} have demonstrated impressive results. Notably, GPT-4 utilized a decomposition model with schema linking and integrated chatbot functionality, leading to a remarkable accuracy rate of 85.3\% as posted on the Spider Text-to-SQL Challenge leaderboard, which currently stands as the highest reported performance.

In 2023, ChatGPT is considered one of the most popular natural language processing tools due to its ability in both language and code generation. Researchers have also assessed the text-to-SQL capability of ChatGPT through zero-shot performance evaluation \cite{Liu2023ACapability}. In this experiment, ChatGPT demonstrated impressive performance without the use of specific training data. It is believed that the text-to-SQL ability of ChatGPT can be further improved through refinement of its training strategy.




\

\section{Conclusion}

In this paper, we discussed recent advances in NLP and DW as an integrated solution for information retrieving from OLTP or DW using human speech. 

We have demonstrated a comprehensive workflow that encompasses text processing and the development of NLP models for converting neural language queries into SQL statements. Our approach involved the use of the pre-trained T5 language model, the publicly available Spider dataset, and customized datasets based on the database structure with the partner company.  The exact match rates for accurately generating SQL for the OLTP and the data warehouse from natural language queries are 73\% and 85\% respectively.

The research demonstrates, through the implementation of the models developed in conjunction with the other work on the chatbot and data warehouse in the research project, that a useful natural language to SQL could be developed rapidly with basic computer resources. More work could be done on using the same approach on a more powerful pre-train language model. Also, research on the generation of the set of training queries based on the database schema would further improve the speed of the development efforts and the deployment of the model to other database environments.

\section*{Acknowledgment}

We acknowledge and thank Diomari Fortes of Okanagan College for his work in building a testing data warehouse and gathering research as a student and former member of our research team. Students from two capstone project teams at Langara College have also contributed to the research and development of an early prototype of the NL to SQL model. For that, we thank you very much.

\bibliographystyle{IEEEtran}
\balance

\bibliography{exportDW4.bib}

\begin{thebibliography}{10}
\providecommand{\url}[1]{#1}
\csname url@samestyle\endcsname
\providecommand{\newblock}{\relax}
\providecommand{\bibinfo}[2]{#2}
\providecommand{\BIBentrySTDinterwordspacing}{\spaceskip=0pt\relax}
\providecommand{\BIBentryALTinterwordstretchfactor}{4}
\providecommand{\BIBentryALTinterwordspacing}{\spaceskip=\fontdimen2\font plus
\BIBentryALTinterwordstretchfactor\fontdimen3\font minus \fontdimen4\font\relax}
\providecommand{\BIBforeignlanguage}[2]{{%
\expandafter\ifx\csname l@#1\endcsname\relax
\typeout{** WARNING: IEEEtran.bst: No hyphenation pattern has been}%
\typeout{** loaded for the language `#1'. Using the pattern for}%
\typeout{** the default language instead.}%
\else
\language=\csname l@#1\endcsname
\fi
#2}}
\providecommand{\BIBdecl}{\relax}
\BIBdecl

\bibitem{RashedMohassel2014AInfrastructure}
R.~Rashed~Mohassel, A.~Fung, F.~Mohammadi, and K.~Raahemifar, ``{A survey on Advanced Metering Infrastructure},'' \emph{International Journal of Electrical Power and Energy Systems}, vol.~63, pp. 473--484, 2014.

\bibitem{Wong2021ASystems}
A.~Wong, D.~Joiner, C.~Chiu, M.~Elsayed, K.~Pereira, Y.~Khmelevsky, and J.~Mahony, ``{A Survey of Natural Language Processing Implementation for Data Query Systems},'' in \emph{RASSE 2021 - IEEE International Conference on Recent Advances in Systems Science and Engineering, Proceedings}, 2021.

\bibitem{Affolter2019ADatabases}
K.~Affolter, K.~Stockinger, and A.~Bernstein, ``{A comparative survey of recent natural language interfaces for databases},'' \emph{VLDB Journal}, vol.~28, no.~5, pp. 793--819, 10 2019.

\bibitem{Li2023RESDSQL:Text-to-SQL}
\BIBentryALTinterwordspacing
H.~Li, J.~Zhang, C.~Li, and H.~Chen, ``{RESDSQL: Decoupling Schema Linking and Skeleton Parsing for Text-to-SQL},'' in \emph{Proceedings of the Thirty-Seventh AAAI Conference on Artificial Intelligence (AAAI)}, 2 2023. [Online]. Available: \url{http://arxiv.org/abs/2302.05965}
\BIBentrySTDinterwordspacing

\bibitem{Li2023Graphix-T5:Parsing}
\BIBentryALTinterwordspacing
J.~Li, B.~Hui, R.~Cheng, B.~Qin, C.~Ma, N.~Huo, F.~Huang, W.~Du, L.~Si, and Y.~Li, ``{Graphix-T5: Mixing Pre-Trained Transformers with Graph-Aware Layers for Text-to-SQL Parsing},'' 1 2023. [Online]. Available: \url{http://arxiv.org/abs/2301.07507}
\BIBentrySTDinterwordspacing

\bibitem{Pourreza2023DIN-SQL:Self-Correction}
\BIBentryALTinterwordspacing
M.~Pourreza and D.~Rafiei, ``{DIN-SQL: Decomposed In-Context Learning of Text-to-SQL with Self-Correction},'' \emph{arXiv preprint arXiv:2304.11015}, 4 2023. [Online]. Available: \url{http://arxiv.org/abs/2304.11015}
\BIBentrySTDinterwordspacing

\bibitem{Joiner2022DWStudy}
D.~Joiner, M.~Clement, S.~Chan, K.~Pereira, A.~Wong, Y.~Khmelevsky, J.~Mahony, and M.~Ferri, ``{DW vs OLTP Performance Optimization in the Cloud on PostgreSQL (A Case Study)},'' in \emph{RASSE 2022 - IEEE International Conference on Recent Advances in Systems Science and Engineering, Symposium Proceedings}, 2022.

\bibitem{Uma2019FormationNLP}
M.~Uma, V.~Sneha, G.~Sneha, J.~Bhuvana, and B.~Bharathi, ``{Formation of SQL from Natural Language Queryusing NLP},'' in \emph{2019 International Conference on Computational Intelligence in Data Science (ICCIDS)}.\hskip 1em plus 0.5em minus 0.4em\relax IEEE, 2019.

\bibitem{Deng2022RecentExpect}
\BIBentryALTinterwordspacing
N.~Deng, Y.~Chen, and Y.~Zhang, ``{Recent Advances in Text-to-SQL: A Survey of What We Have and What We Expect},'' \emph{arXiv preprint arXiv:2208.10099.}, 8 2022. [Online]. Available: \url{http://arxiv.org/abs/2208.10099}
\BIBentrySTDinterwordspacing

\bibitem{Yu2018Spider:Task}
\BIBentryALTinterwordspacing
T.~Yu, R.~Zhang, K.~Yang, M.~Yasunaga, D.~Wang, Z.~Li, J.~Ma, I.~Li, Q.~Yao, S.~Roman, Z.~Zhang, and D.~Radev, ``{Spider: A Large-Scale Human-Labeled Dataset for Complex and Cross-Domain Semantic Parsing and Text-to-SQL Task},'' \emph{arXiv preprint arXiv:1809.08887}, 9 2018. [Online]. Available: \url{http://arxiv.org/abs/1809.08887}
\BIBentrySTDinterwordspacing

\bibitem{Li2020AModel}
Q.~Li, L.~Li, Q.~Li, and J.~Zhong, ``{A Comprehensive Exploration on Spider with Fuzzy Decision Text-to-SQL Model},'' \emph{IEEE Transactions on Industrial Informatics}, vol.~16, no.~4, pp. 2542--2550, 4 2020.

\bibitem{Guo2019ContentGeneration}
\BIBentryALTinterwordspacing
T.~Guo and H.~Gao, ``{Content Enhanced BERT-based Text-to-SQL Generation},'' \emph{arXiv preprint arXiv:1910.07179}, 10 2019. [Online]. Available: \url{http://arxiv.org/abs/1910.07179}
\BIBentrySTDinterwordspacing

\bibitem{Zhong2017Seq2SQL:Learning}
\BIBentryALTinterwordspacing
V.~Zhong, C.~Xiong, and R.~Socher, ``{Seq2SQL: Generating Structured Queries from Natural Language using Reinforcement Learning},'' \emph{arXiv preprint arXiv:1709.00103.}, 8 2017. [Online]. Available: \url{http://arxiv.org/abs/1709.00103}
\BIBentrySTDinterwordspacing

\bibitem{Mellah2021SQLNetworks}
Y.~Mellah, E.~H. Ettifouri, A.~Rhouati, W.~Dahhane, T.~Bouchentouf, and M.~G. Belkasmi, ``{SQL Generation from Natural Language Using Supervised Learning and Recurrent Neural Networks},'' in \emph{Lecture Notes in Networks and Systems}.\hskip 1em plus 0.5em minus 0.4em\relax Springer, 2021, vol. 144, pp. 175--183.

\bibitem{Bogin2019RepresentingParsing}
\BIBentryALTinterwordspacing
B.~Bogin, M.~Gardner, and J.~Berant, ``{Representing Schema Structure with Graph Neural Networks for Text-to-SQL Parsing},'' \emph{arXiv preprint arXiv:1905.06241.}, 5 2019. [Online]. Available: \url{http://arxiv.org/abs/1905.06241}
\BIBentrySTDinterwordspacing

\bibitem{Chen2021ShadowGNN:Parser}
\BIBentryALTinterwordspacing
Z.~Chen, L.~Chen, Y.~Zhao, R.~Cao, Z.~Xu, S.~Zhu, and K.~Yu, ``{ShadowGNN: Graph Projection Neural Network for Text-to-SQL Parser},'' \emph{arXiv preprint arXiv:2104.04689}, 4 2021. [Online]. Available: \url{http://arxiv.org/abs/2104.04689}
\BIBentrySTDinterwordspacing

\bibitem{Hui2022S2SQL:Parsers}
\BIBentryALTinterwordspacing
B.~Hui, R.~Geng, L.~Wang, B.~Qin, B.~Li, J.~Sun, and Y.~Li, ``{S2SQL: Injecting Syntax to Question-Schema Interaction Graph Encoder for Text-to-SQL Parsers},'' \emph{arXiv preprint arXiv:2203.06958.}, 3 2022. [Online]. Available: \url{http://arxiv.org/abs/2203.06958}
\BIBentrySTDinterwordspacing

\bibitem{Cai2021SADGA:Text-to-SQL}
\BIBentryALTinterwordspacing
R.~Cai, J.~Yuan, B.~Xu, and Z.~Hao, ``{SADGA: Structure-Aware Dual Graph Aggregation Network for Text-to-SQL},'' \emph{Neural Information Processing Systems, 34, 7664-7676.}, vol.~34, pp. 7664--7676, 10 2021. [Online]. Available: \url{http://arxiv.org/abs/2111.00653}
\BIBentrySTDinterwordspacing

\bibitem{Wang2019RAT-SQL:Parsers}
\BIBentryALTinterwordspacing
B.~Wang, R.~Shin, X.~Liu, O.~Polozov, and M.~Richardson, ``{RAT-SQL: Relation-Aware Schema Encoding and Linking for Text-to-SQL Parsers},'' \emph{arXiv preprint arXiv:1911.04942}, 11 2019. [Online]. Available: \url{http://arxiv.org/abs/1911.04942}
\BIBentrySTDinterwordspacing

\bibitem{Pennington2014GloVe:Representation}
\BIBentryALTinterwordspacing
J.~Pennington, R.~Socher, and C.~D. Manning, ``{GloVe: Global Vectors for Word Representation},'' in \emph{Proceedings of the 2014 conference on empirical methods in natural language processing (EMNLP)}, 2014, pp. 1532--1543. [Online]. Available: \url{http://nlp.}
\BIBentrySTDinterwordspacing

\bibitem{Ferreira2020EvaluatingSystems}
S.~Ferreira, G.~Leit{\~{a}}o, I.~Silva, A.~Martins, and P.~Ferrari, ``{Evaluating human-machine translation with attention mechanisms for industry 4.0 environment SQL-based systems},'' in \emph{IEEE International Workshop on Metrology for Industry 4.0 {\&} IoT}, 2020, pp. 229--234.

\bibitem{Vathsala2021NLP2SQLLearning}
H.~Vathsala and S.~Koolagudi, ``{NLP2SQL Using Semi-supervised Learning},'' in \emph{Advanced Computing: 10th International Conference, IACC 2020, Panaji, Goa, India, December 5–6, 2020 , Revised Selected Papers, Part I 10}.\hskip 1em plus 0.5em minus 0.4em\relax Springer, 2021, pp. 288--299.

\bibitem{Xu2017SQLNet:Learning}
\BIBentryALTinterwordspacing
X.~Xu, C.~Liu, and D.~Song, ``{SQLNet: Generating Structured Queries From Natural Language Without Reinforcement Learning},'' \emph{arXiv preprint arXiv:1711.04436}, 11 2017. [Online]. Available: \url{http://arxiv.org/abs/1711.04436}
\BIBentrySTDinterwordspacing

\bibitem{Wang2022Pre-TrainedApplications}
H.~Wang, J.~Li, H.~Wu, E.~Hovy, and Y.~Sun, ``{Pre-Trained Language Models and Their Applications},'' \emph{Engineering}, 9 2022.

\bibitem{Brunner2021ValueNet:Information}
\BIBentryALTinterwordspacing
U.~Brunner and K.~Stockinger, ``{ValueNet: A Natural Language-to-SQL System that Learns from Database Information},'' in \emph{2021 IEEE 37th International Conference on Data Engineering (ICDE}, 2021, pp. 2177--2182. [Online]. Available: \url{https://yale-lily.github.io/spider}
\BIBentrySTDinterwordspacing

\bibitem{Hwang2019AContextualization}
\BIBentryALTinterwordspacing
W.~Hwang, J.~Yim, S.~Park, and M.~Seo, ``{A Comprehensive Exploration on WikiSQL with Table-Aware Word Contextualization},'' \emph{arXiv preprint arXiv:1902.01069}, 2 2019. [Online]. Available: \url{http://arxiv.org/abs/1902.01069}
\BIBentrySTDinterwordspacing

\bibitem{Choi2020RYANSQL:Databases}
\BIBentryALTinterwordspacing
D.~Choi, M.~C. Shin, E.~Kim, and D.~R. Shin, ``{RYANSQL: Recursively Applying Sketch-based Slot Fillings for Complex Text-to-SQL in Cross-Domain Databases},'' \emph{Computational Linguistics}, vol.~47, no.~2, pp. 309--332, 4 2020. [Online]. Available: \url{http://arxiv.org/abs/2004.03125}
\BIBentrySTDinterwordspacing

\bibitem{Guo2019TowardsRepresentation}
\BIBentryALTinterwordspacing
J.~Guo, Z.~Zhan, Y.~Gao, Y.~Xiao, J.-G. Lou, T.~Liu, and D.~Zhang, ``{Towards Complex Text-to-SQL in Cross-Domain Database with Intermediate Representation},'' \emph{arXiv preprint arXiv:1905.08205}, 5 2019. [Online]. Available: \url{http://arxiv.org/abs/1905.08205}
\BIBentrySTDinterwordspacing

\bibitem{Cao2021LGESQL:Relations}
\BIBentryALTinterwordspacing
R.~Cao, L.~Chen, Z.~Chen, Y.~Zhao, S.~Zhu, and K.~Yu, ``{LGESQL: Line Graph Enhanced Text-to-SQL Model with Mixed Local and Non-Local Relations},'' \emph{arXiv preprint arXiv:2106.01093.}, 6 2021. [Online]. Available: \url{http://arxiv.org/abs/2106.01093}
\BIBentrySTDinterwordspacing

\bibitem{Elazar2021MeasuringModels}
\BIBentryALTinterwordspacing
Y.~Elazar, N.~Kassner, S.~Ravfogel, A.~Ravichander, E.~Hovy, H.~Sch{\"{u}}tze, S.~Sch{\"{u}}tze, and Y.~Goldberg, ``{Measuring and Improving Consistency in Pretrained Language Models},'' \emph{Transactions of the Association for Computational Linguistics}, vol.~9, pp. 1012--1031, 2021. [Online]. Available: \url{https://doi.org/10.1162/tacl}
\BIBentrySTDinterwordspacing

\bibitem{Devlin2018BERT:Understanding}
\BIBentryALTinterwordspacing
J.~Devlin, M.-W. Chang, K.~Lee, and K.~Toutanova, ``{BERT: Pre-training of Deep Bidirectional Transformers for Language Understanding},'' \emph{arXiv preprint arXiv:1810.04805.}, 10 2018. [Online]. Available: \url{http://arxiv.org/abs/1810.04805}
\BIBentrySTDinterwordspacing

\bibitem{Zhang2019Editing-BasedQuestions}
R.~Zhang, T.~Yu, H.~Y. Er, S.~Shim, E.~Xue, X.~V. Lin, T.~Shi, C.~Xiong, R.~Sovher, and D.~Radev, ``{Editing-Based SQL Query Generation for Cross-Domain Context-Dependent Questions},'' \emph{arXiv preprint arXiv:1909.00786.}, 2019.

\bibitem{Ma2020MentionGeneration}
\BIBentryALTinterwordspacing
J.~Ma, Z.~Yan, S.~Pang, Y.~Zhang, and J.~Shen, ``{Mention Extraction and Linking for SQL Query Generation},'' \emph{arXiv preprint arXiv:2012.10074}, 12 2020. [Online]. Available: \url{http://arxiv.org/abs/2012.10074 http://dx.doi.org/10.18653/v1/2020.emnlp-main.563}
\BIBentrySTDinterwordspacing

\bibitem{Lin2020BridgingParsing}
\BIBentryALTinterwordspacing
X.~V. Lin, R.~Socher, and C.~Xiong, ``{Bridging Textual and Tabular Data for Cross-Domain Text-to-SQL Semantic Parsing},'' \emph{arXiv preprint arXiv:2012.12627}, 12 2020. [Online]. Available: \url{http://arxiv.org/abs/2012.12627}
\BIBentrySTDinterwordspacing

\bibitem{Shaw2020CompositionalBoth}
\BIBentryALTinterwordspacing
P.~Shaw, M.-W. Chang, P.~Pasupat, and K.~Toutanova, ``{Compositional Generalization and Natural Language Variation: Can a Semantic Parsing Approach Handle Both?}'' \emph{arXiv preprint arXiv:2010.12725.}, 10 2020. [Online]. Available: \url{http://arxiv.org/abs/2010.12725}
\BIBentrySTDinterwordspacing

\bibitem{Raffel2019ExploringTransformer}
\BIBentryALTinterwordspacing
C.~Raffel, N.~Shazeer, A.~Roberts, K.~Lee, S.~Narang, M.~Matena, Y.~Zhou, W.~Li, and P.~J. Liu, ``{Exploring the Limits of Transfer Learning with a Unified Text-to-Text Transformer},'' \emph{The Journal of Machine Learning Research}, vol.~21, no.~1, pp. 5485--5551, 10 2019. [Online]. Available: \url{http://arxiv.org/abs/1910.10683}
\BIBentrySTDinterwordspacing

\bibitem{Xie2022UnifiedSKG:Models}
\BIBentryALTinterwordspacing
T.~Xie, C.~H. Wu, P.~Shi, R.~Zhong, T.~Scholak, M.~Yasunaga, C.-S. Wu, M.~Zhong, P.~Yin, S.~I. Wang, V.~Zhong, B.~Wang, C.~Li, C.~Boyle, A.~Ni, Z.~Yao, D.~Radev, C.~Xiong, L.~Kong, R.~Zhang, N.~A. Smith, L.~Zettlemoyer, and T.~Yu, ``{UnifiedSKG: Unifying and Multi-Tasking Structured Knowledge Grounding with Text-to-Text Language Models},'' \emph{arXiv preprint arXiv:2201.05966}, 1 2022. [Online]. Available: \url{http://arxiv.org/abs/2201.05966}
\BIBentrySTDinterwordspacing

\bibitem{Scholak2021PICARD:Models}
\BIBentryALTinterwordspacing
T.~Scholak, N.~Schucher, and D.~Bahdanau, ``{PICARD: Parsing Incrementally for Constrained Auto-Regressive Decoding from Language Models},'' \emph{arXiv preprint arXiv:2109.05093}, 9 2021. [Online]. Available: \url{http://arxiv.org/abs/2109.05093}
\BIBentrySTDinterwordspacing

\bibitem{Qi2022RASAT:Text-to-SQL}
\BIBentryALTinterwordspacing
J.~Qi, J.~Tang, Z.~He, X.~Wan, Y.~Cheng, C.~Zhou, X.~Wang, Q.~Zhang, and Z.~Lin, ``{RASAT: Integrating Relational Structures into Pretrained Seq2Seq Model for Text-to-SQL},'' \emph{arXiv preprint arXiv:2205.06983.}, 5 2022. [Online]. Available: \url{http://arxiv.org/abs/2205.06983}
\BIBentrySTDinterwordspacing

\bibitem{Zhao2022ImportanceParsing}
\BIBentryALTinterwordspacing
Y.~Zhao, J.~Jiang, Y.~Hu, W.~Lan, H.~Zhu, A.~Chauhan, A.~Li, L.~Pan, J.~Wang, C.-W. Hang, S.~Zhang, M.~Dong, J.~Lilien, P.~Ng, Z.~Wang, V.~Castelli, and B.~Xiang, ``{Importance of Synthesizing High-quality Data for Text-to-SQL Parsing},'' \emph{arXiv preprint arXiv:2212.08785}, 12 2022. [Online]. Available: \url{http://arxiv.org/abs/2212.08785}
\BIBentrySTDinterwordspacing

\bibitem{Zeng2022N-BestSystems}
\BIBentryALTinterwordspacing
L.~Zeng, S.~H.~K. Parthasarathi, and D.~Hakkani-Tur, ``{N-Best Hypotheses Reranking for Text-To-SQL Systems},'' in \emph{IEEE Spoken Language Technology Workshop (SLT)}, 10 2022, pp. 663--670. [Online]. Available: \url{http://arxiv.org/abs/2210.10668}
\BIBentrySTDinterwordspacing

\bibitem{Liu2022AugmentingSelf-Play}
\BIBentryALTinterwordspacing
Q.~Liu, Z.~Ye, T.~Yu, P.~Blunsom, and L.~Song, ``{Augmenting Multi-Turn Text-to-SQL Datasets with Self-Play},'' \emph{arXiv preprint arXiv:2210.12096}, 10 2022. [Online]. Available: \url{http://arxiv.org/abs/2210.12096}
\BIBentrySTDinterwordspacing

\bibitem{Liu2023ACapability}
\BIBentryALTinterwordspacing
A.~Liu, X.~Hu, L.~Wen, and P.~S. Yu, ``{A comprehensive evaluation of ChatGPT's zero-shot Text-to-SQL capability},'' \emph{arXiv preprint arXiv:2303.13547.}, 3 2023. [Online]. Available: \url{http://arxiv.org/abs/2303.13547}
\BIBentrySTDinterwordspacing

\end{thebibliography}
\end{document}